\documentclass[showpacs,aps,prx,twocolumn]{revtex4-1}
\usepackage{amsmath}
\usepackage{amssymb}
\usepackage{graphicx}
\usepackage{ulem}
\usepackage{color,xcolor}

\newcommand{\Tr}{\mathrm{Tr}}
\newcommand{\sinc}{\mathrm{sinc}}

\newcommand{\ele}{{\mathcal L}}

\begin{document}

\title{Microscopic description for the emergence of collective dissipation in extended quantum systems}
\author{Fernando Galve$^1$, Antonio Mandarino$^2$, Matteo G. A. Paris$^{2}$,
Claudia Benedetti$^2$,  Roberta Zambrini$^1$}
\affiliation{$^1$Instituto de F\'isica Interdisciplinar y Sistemas Complejos IFISC (CSIC-UIB), Campus Universitat Illes Balears,
E-07122 Palma de Mallorca, Spain}
\affiliation{$^2$Quantum Technology Lab, Dipartimento di Fisica, Universit{\`a} degli Studi di Milano, I-20133 Milan, Italy }
\date{\today}

\begin{abstract}
Practical implementations of quantum technology are limited by unavoidable effects of decoherence and dissipation. 
With achieved experimental control for individual atoms and photons, more complex platforms composed by several units
can be assembled enabling distinctive forms of dissipation and decoherence,
in independent heat baths or collectively into a common bath, with dramatic consequences for the preservation of 
quantum coherence.
The cross-over between these two regimes has been widely attributed in the literature to the system units being farther apart than the bath's 
correlation length. Starting from a microscopic model of a structured environment (a crystal) sensed
by two bosonic probes, here we show the failure of such conceptual relation, and identify the exact
physical mechanism underlying this cross-over, displaying a sharp contrast between dephasing and dissipative baths.
Depending on the frequency of the system and, crucially, on its orientation with respect to the crystal axes, collective 
dissipation becomes possible for very large distances between probes, opening new avenues to deal with decoherence in phononic baths. 
\end{abstract}

\maketitle

Models for quantum dissipation address the interaction
of a quantum system with bosonic, fermionic or other kinds of 
environments, where the relevant information about the microscopic structure 
of the environment is encoded in its spectral density \cite{weiss,breuer,blattNoise,eisert}. 
On the other hand, further information is required to properly describe 
spatially extended multipartite systems: an often used generalization is the independent
dissipation of the system's components into separate baths (SB),
leading to complete erasure of quantum correlations \cite{weiss,breuer}. Also,
collective or spatially symmetric decoherence into a common bath (CB) 
\cite{palma,rivasmuller,rasetti,duan,white00,dicke,NSS,lidar14,CBent,zhao03,benatti03,pb04,aguado08,paz08,sync,CBcomp,lidar98,bacon00,zanardi,metrology1} 
has been proposed as an alternative scenario
in the limit of small system size (or components separation) in comparison with 
environment correlation length or with radiating atoms' transition wave-length \cite{breuer,palma,rivasmuller}.
A CB opens up outstanding possibilities like superradiance \cite{breuer,dicke}, 
superdecoherence \cite{palma}, and decoherence free/noiseless subspaces \cite{NSS,lidar14}, allowing the
preservation and also creation of entanglement \cite{CBent,zhao03,benatti03,pb04,aguado08,paz08}, the emergence of collective synchronization
\cite{sync},
with potential applications in quantum computation 
\cite{rasetti,duan,white00,CBcomp,lidar98,bacon00,zanardi} and metrology \cite{metrology1}.

Besides artificial methods to engineer collective dissipation mechanisms \cite{blatt,cirac}, the cross-over between CB to SB
can naturally arise in structured environments. The still open and fundamental question is: {\it how small needs to be a multipartite system to dissipate 
 collectively?}
The CB/SB cross-over when increasing the size of spatially extended systems  has been phenomenologically modeled
in the last decade yielding a smooth change and, generally, assuming isotropic dispersion relations of bosonic environments 
\cite{kohler2006,fisher2009,Klesse,JeskeCole}
(like it happens for electromagnetic radiation in free-space  \cite{breuer,rivasmuller}). 
Assuming a distance dependent transition from collective to independent dissipation, important predictions have been reported in the context of quantum error
correction \cite{preskill}, in the dynamics of photosynthetic complexes
\cite{aspuru,olaya,thorwart} and in quantum metrology \cite{metrology2}. 
Even if a microscopic derivation of the CB/SB cross-over is {\it still missing} in spatially structured environments, 
 it is usually argued that a common environmental medium with significant  spatial correlations up to distances $\xi_c$
will produce both damping for each system unit and a {\it cross-damping} among them: a collective dissipation is therefore generally 
associated to systems smaller than the correlation length  $\xi_c$, while units far away will be damped independently in SB. 
Here we are going to show the failure of this prediction for a large class of energy-matter exchange dissipation models, particularizing
to a specific microscopic model to clarify and illustrate several details: a phonon bath in
a crystal probed at different spatial locations. We address the cross-over from CB to SB
in detail, providing a physical ground for the description of intermediate regimes, and assessing the role played by geometric factors, 
spatial extension of the system-probe contact and bath correlations.
Our model allows to clarify several issues including: a) 
why when increasing the system size in 1D environments
\cite{kohler2006,fisher2009,JeskeCole,morigi} there is no asymptotic
interpolation between CB and SB, but a periodic cross-over; 
b) why choosing an isotropic environmental dispersion relation 
will always lead to distance-decaying cross-damping, c) why 
anisotropic dispersion relations (like those in real 
crystals with symmetries) can lead to surprising
effects like CB at large distances, also showing d) that in general the 
correlation length is {\it not} related to the CB/SB transition. We further e)
give a simple intuitive picture of how a bath's frequency cutoff appears naturally
from the fact that the system's quantum units have a finite spatial extent, and f)
we show how the presence of static disorder favours SB dissipation.

For clarity we introduce next a particular model displaying all the phenomenology, and 
leave the discussion on the generality of these effects to the last section.


\section{Environment induced cross-talk} 
We consider a $D-$dimensional periodic crystal, in the same spirit that led Rubin \cite{Rubin} 
to introduce a linear harmonic chain as a microscopic model of 
an Ohmic bosonic bath \cite{morigi,vasile}. 
This model allows to model spatially correlated dissipation and
provides a common ground to assess the role of
different crystal dimensionality D and geometries, including spatial disorder effects,
either for point-like and for non-local system-bath interactions. 
%
\begin{figure}[b!]
\includegraphics[width=\columnwidth]{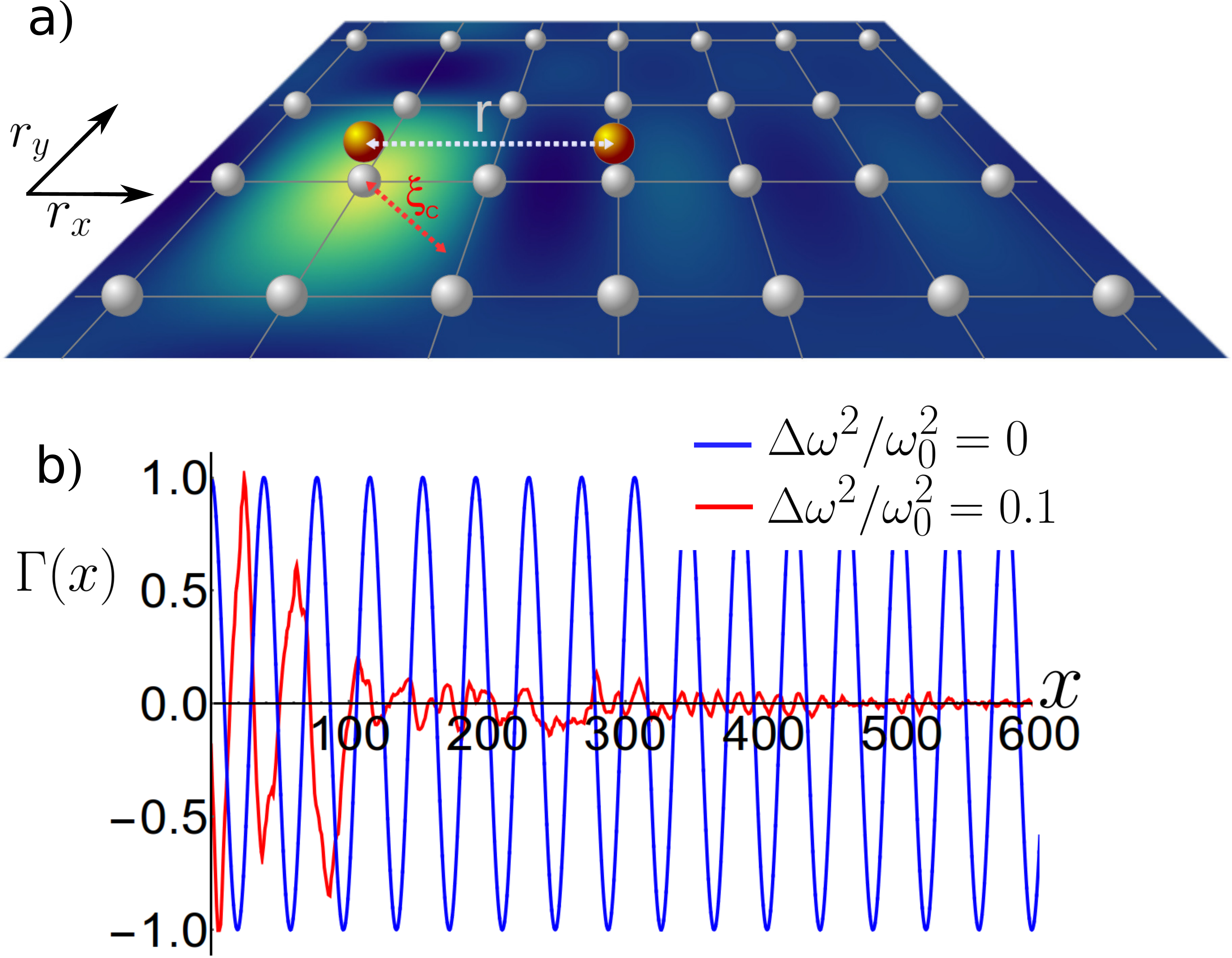}
\caption{(a) Sketch of a 2D crystal and two locally attached probes, at distance $\vec{r}$. We pictorially plot the bath's correlation
(of spatial extent $\xi_c$) centered in one  probe (see also Fig. \ref{fig4}c).
(b) Cross-talk for the 1D periodic and disordered environment as a function of the probes distance.
Added random noise in the onsite potential with amplitude $\Delta\omega^2=0,0.1$, $\omega_0=1$, $g=3/4$, probe frequency $\Omega$ resonant with $k_\Omega=0.164$. 
The normalized cross-talk in absence of noise is $\Gamma(x)\hat{=}\Gamma_{13}^{(1)}(x)/\Gamma_{13}^{(1)}(0)=cos(k_\Omega x)$ while in presence of disorder it is  position dependent due to lack of
 translational invariance, $\Gamma_{n_0}(x)\hat{=}\Gamma_{13}^{(1)}(n_0,n_0+x)/\Gamma_{13}^{(1)}(n_0,n_0)$. We present $\Gamma_{n_0}(x)$ for an arbitrary $n_0$ and a given noise realization, with $x\in[0,600]$ and a
we have used a finite harmonic chain of 2500 oscillators.
} 
\label{fig1}
\end{figure}
The $D$-dimensional crystal consists on an infinite collection of harmonically coupled masses ($\hbar=m=1$) with on-site harmonic potential    
of frequency $\omega_0$ (see Fig.~\ref{fig1}a for a representation for $D=2$). We focus for the sake of simplicity in oscillations in one direction
corresponding to one phonon polarization  (see Appendix~\ref{AppendixA}).
The dissipative system consists of two probes whose distance $\vec{r}$ can be tuned, namely
two uncoupled harmonic oscillators of frequency $\Omega$ 
{\it weakly} interacting with the crystal. We start considering point-like contacts 
at two different spatial locations $\vec{n}$ and $\vec{n'}=\vec{n}+\vec{r}$.

\par
The master equation for the reduced density matrix of the 
two probes may be obtained within the 
Born-Markov approximation and assuming the environment in a 
Gibbs state at temperature $T$ \cite{breuer}
\begin{equation}
\dot{ \rho}(t) = -i \left[ \rho, \tilde{H}_S \right]+ \sum_{j ,l=1}^4 
\Gamma^{(D)}_{j l}(\vec{r},t) ( F_j \rho F_l^\dag - \frac12 \{F_l^\dag F_j,
\rho\})\,, 
\label{lindblad}
\end{equation} 
where $\tilde{H}_S=H_S+H_{LS}$, with $H_S$ and $H_{LS}$ the system Hamiltonian and bath's Lamb-shift (see eq. (\ref{fullME}) in Appendix~\ref{AppendixA}).
The $F_j = \{ a_1, a_1^\dag, a_2, a_2^\dag \} $ are the 
annihilation (creation) operators of each probe and 
$\Gamma^{(D)}_{j l}$ are the corresponding 
damping coefficients (the superscript refers to the 
dimensionality of the crystal), depending only on the distance $\vec{r}$ owing to environment 
translational invariance. Self-damping of each oscillator ($j=l$) 
and cross terms ($|j-l|=2$) characterize the dissipation with 
 $$\Gamma^{(D)}_{j l}(\vec{r}, t)= \lambda^2(2\pi)^{-D}\int_{-\pi}^{\pi} d^D\vec{k}\,  
C_{j l}(\vec{k},\vec{r},t)/(2\Omega\ \omega_{\vec{k}})$$
and non-vanishing terms
\begin{eqnarray}
C_{11} &=&  [n(\vec{k}) + 1]\sin[(\omega_{\vec{k}}-
\Omega) t]/[\omega_{\vec{k}}- \Omega]=C_{33}\nonumber\\ 
C_{22} &=&  n(\vec{k})\, \sin[(\omega_{\vec{k}}-
\Omega) t]/[\omega_{\vec{k}}- \Omega]=C_{44}\nonumber\\ 
C_{13} &=& C_{11} \cos( \vec{k} \cdot \vec{r})\, , \, C_{24} = C_{22} \cos( \vec{k} \cdot \vec{r})\nonumber. 
\end{eqnarray}
 For a
bath at zero temperature, the only nonzero coefficients are the the
self-damping $\Gamma_{11}=\Gamma_{33}$ and cross term $\Gamma_{13}$.

A crucial point is that if the two probes are attached
to a common environmental point (CB case), i.e. $H_{SB}=\lambda(a_1+a_2) 
A_{\vec{n}}^\dagger+h.c.$, we have $\Gamma_{11}=\Gamma_{13}$ and
$\Gamma_{22}=\Gamma_{24}$ \cite{JeskeCole}, whereas for probes attached to two  independent 
environments (SB case) we would have 
$\Gamma_{13}=\Gamma_{24}=0$, i.e. no cross terms. The cross-over between CB and SB regimes depending on the probes distance 
can now be derived  from this microscopic model without further assumptions.
For long times, when the weak dissipation becomes important, only a family of resonant momentum  crystal phonons are relevant,
such that $\omega(\vec{k}_\Omega )=\Omega$. This condition identifies the manifold of phonons 
mediating an eventual cross-talk between the oscillators. The
dependence on probes distance at $T=0$ and long times is then
\begin{equation}
\Gamma^{(D)}_{13}(\vec{r})=\frac{\lambda^2}{2\Omega^2(2\pi)^D}\int\!\!d^D\vec{k}\cos(\vec{k}\vec{r}).
\delta(\omega_{\vec{k}}-\Omega)
\label{0Tcross}
\end{equation}
Note that in the weak damping regime we are considering here, the dissipation rate $\lambda^2$ is much smaller than the frequencies of the problem, which guarantees
that at times where the quantum units start `feeling' dissipation, the sinc function is well approximated by a delta.

\section{The exceptional 1D case and disorder effects} Notice that an immediate consequence, previously observed in \cite{kohler2006,JeskeCole,fisher2009}, but scarcely 
commented upon, is that for 1D homogeneous environments, irrespective of the dispersion relation, we have $\Gamma^{(1)}_{13}(x)\propto \cos(k_\Omega x)$, since the frequency
resonance constraint exhausts all freedom in choosing the crystal momenta in eq. (\ref{0Tcross}). 
This means that two probes will experience collective dissipation not only when attached to the same point of the environment but also 
when at the anti-nodes of the resonant mode \cite{morigi}. 
In this case the relative 
position or the center of mass of the pair is shielded from decoherence, allowing to preserve  entanglement among the probes at large distances.
Indeed the surprising results is the lack of asymptotic cross-damping decay above any distance, being the cross-over between SB and CB periodically predicted.
Further, if the relative size of cross-damping and self-damping are considered, this result is unchanged when increasing the 
temperature of the thermal bath (this is due to $n(\vec{k})$ factoring out of the integrals
because it depends only on the frequency).  

The generalization to higher dimensional environment leads to a richer scenario, but before proceeding it is interesting to assess the fragility of this phenomenon in experiments 
considering the effect of {\it static disorder}.
The cross-talk can be understood as the sum of overlaps of resonant crystal normal modes at the probes positions.
The expression (2) obtained when plane waves are the normal modes, can be in general 
expressed as $$\Gamma_{13}^{(D)}(\vec{n},\vec{n}')=(\lambda^2/2\Omega)\int d^D\vec{k}f_{\vec{n},\vec{k}} f_{\vec{n}',\vec{k}}^* \frac{\sin[t(\omega_{\vec{k}}-\Omega)]}{\omega_{\vec{k}}-\Omega}\frac{1}{\omega_{\vec{k}}}$$ where  $f_{\vec{n},\vec{k}}$ is the spatial profile of eigenmode $\vec{k}$,
and now the cross-talk is position dependent ($\Gamma_{13}^{(D)}(\vec{n},\vec{n}')\neq \Gamma_{13}^{(D)}(\vec{r})$).
The presence of disorder, here modeled by inhomogeneity in the local crystal potentials, breaks translational invariance 
and leads to localized waves. As a consequence the cross-talk,
periodic in the homogeneous case, now decays  with the distance at an average rate depending on the degree of disorder,
as shown in Fig.\ref{fig1}b.  
This localization effect \cite{anderson} hinders  the periodic cross-over between CB and SB leading to a spatial decay:
beyond some distance, two independent probes will dissipate into SB.   

\section{Isotropic vs. anisotropic cases} When moving to $D>1$ a common assumption in several phenomenological approaches, either based on spin-boson
\cite{rivasmuller,zanardi,kohler2006,fisher2009,JeskeCole,metrology2}
or boson-boson models \cite{Klesse}, is the
isotropy of the dispersion relation of the environment, i.e. its dependence only on the
modulus $|\vec{k}|$. This is the case for electromagnetic environment \cite{breuer}. The isotropy
of the environment dispersion enables some analytical insight and leads to  a {\it spatially
decaying} cross-talk in the master equation. For $T=0$ and long times the cross-talk dependence on the environment dimension is
\begin{eqnarray}
\Gamma_{13}^{(1)}( r)&\propto&\cos(|k_\Omega|  x)\nonumber\\
\Gamma_{13}^{(2)}(\vec{r})&\propto& J_0(|\vec{k}_\Omega|  r)\nonumber\\
\Gamma_{13}^{(3)}(\vec{r})&\propto& \sinc(|\vec{k}_\Omega|  r)\nonumber
\end{eqnarray}
with $\omega(\vec{k}_\Omega )=\Omega$.

\begin{figure}[t!]
\includegraphics[width=1.05\columnwidth]{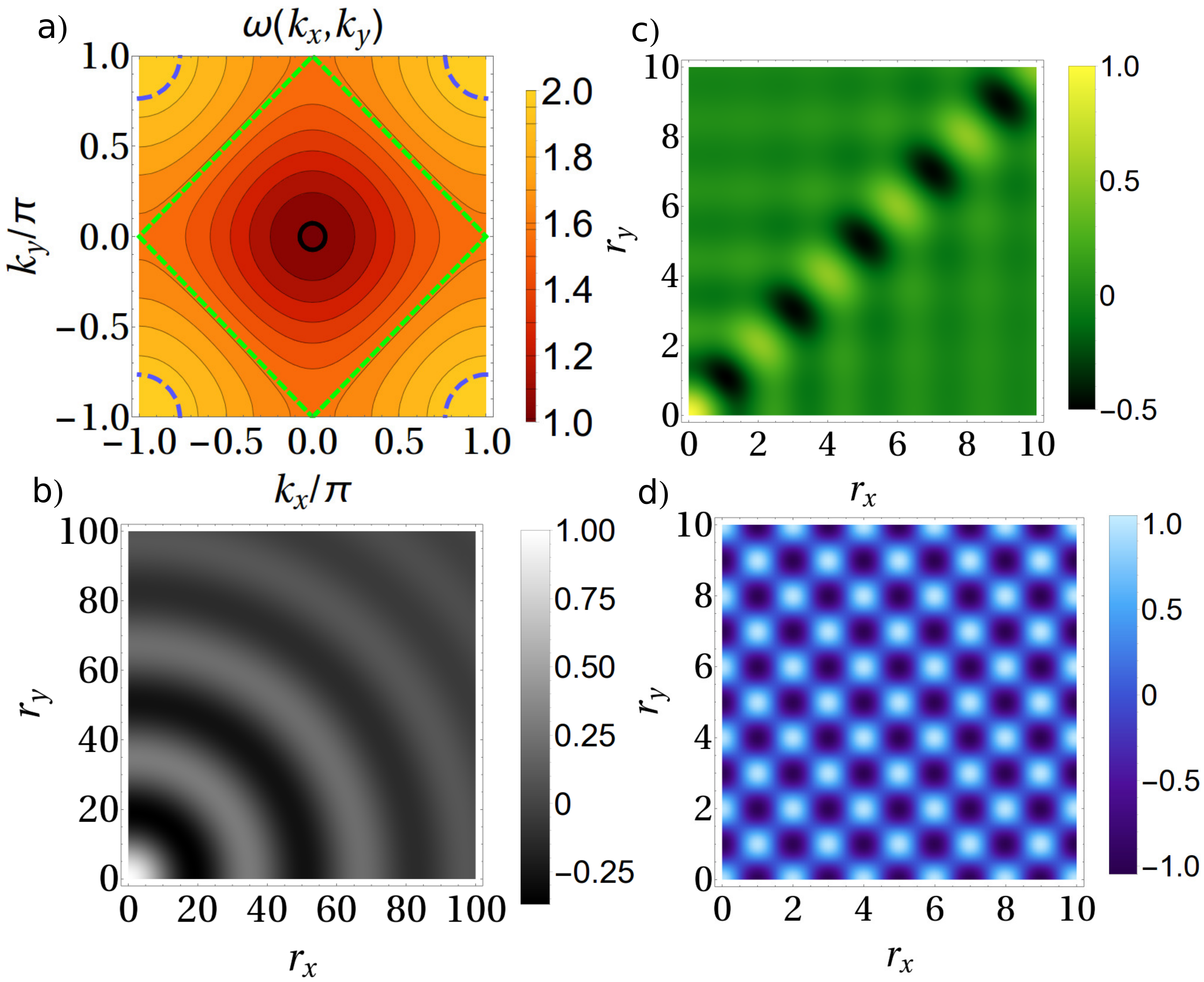}
\caption{
a) 2D dispersion relation in color code with $\omega_0=1$ and $g=3/16$, so that $\omega_{\vec{k}}\in [1,2]$.
Iso-frequency surfaces are shown
for the limiting cases discussed in the text:  
black) $\Omega=1.01$ corresponding to the isotropic case, 
green) $\Omega=\sqrt{5/2}$ and blue) $\Omega=1.95$.
Normalized cross-damping term $\Gamma(r_x,r_y)\hat{=}\Gamma_{13}^{(2)}(\vec{r})/\Gamma_{13}^{(2)}(0)$ for 
b) the isotropic case (low momenta), for 
c) directional non-decay (medium momenta) and 
d) non-decay (high momenta) (see text for details). We plot only one spatial quadrant because of 
the symmetry of the setting.}
\label{fig2}
\end{figure}

On the other hand the dispersion in spatially structured media are typically not isotropic.
In the case of a cubic homogeneous crystal, for instance,  $$\omega_{\vec{k}}=\sqrt{\omega_0^2+4D g(\sin^2\frac{k_x}{2}+
\sin^2\frac{k_y}{2}+...+\sin^2\frac{k_D}{2})}$$ where we recognize the effect of the spatial symmetries (we discuss later the 
triangular case). 
Still the dispersion is approximately isotropic for small momenta (Fig.~\ref{fig2} black circle) 
($\omega_{\vec{k}}\simeq\omega_{|\vec{k}|}=\sqrt{\omega_0^2+D g|\vec{k}|^2}$), and the angular integration yields a function
decaying with the radial distance between probes (Fig.~\ref{fig2}b). Independently on the crystal direction probed by the system components, 
collective dissipation is lost above some distance where the crystal will effectively acts as two SB.

Departure from isotropic dispersion
relations has deep consequences. Although in general there
will be a spatial (non-monotonic) decay of $\Gamma_{13}^{(D)}(\vec{r})$, 
different scenarios may arise like those of Fig.~\ref{fig2}c and \ref{fig2}d. In general the anisotropy of the dispersion will translate into a
sensitivity of the probes dissipation to the crystal geometry.
In Fig.~\ref{fig2}c we observe  for a particular resonance value
$\Omega$   an interference effect resulting
in decay of $\Gamma_{13}^{(d)}(\vec{r})$ along all directions {\it
except} for the lattice diagonals $y=\pm x$ where it does not decay. Indeed
$\vec{k}_\Omega
=\{k_x,\pm (\pi-|k_x|)\}$ yields 
$\Gamma_{13}^{(2)}(\vec{r})\propto (x\sin(\pi x)-y\sin(\pi y))/(x^2-y^2)$, not decaying on the crystal diagonals. 
Strong anisotropy is also displayed in Fig.~\ref{fig2}d, for 
$k_x=\pm\pi$, $k_y=\pm\pi$ and leading to a periodic cross-term $\Gamma_{13}^{(2)}(\vec{r})\propto \cos(\pi r_x) \cos(\pi r_y)$.
Then no asymptotic decay of the cross-damping  with distance occurs and these high frequency probes are able to `resolve' the spatial
structure of the crystal.

Similar results are found in 3D: resonant momenta for a given $\Omega$ will lie in a surface, 
and cross-talk will depend on their interference. 
For isotropic (low momenta) case we have the form $\Gamma_{13}^{(3)}(|\vec{r}|)\propto \sinc(|\vec{k}_\Omega| |\vec{r}|)$, while for high momentum we have a similar 'egg-crate' 
in 3D $\Gamma_{13}^{(3)}(\vec{r})\propto \cos(\pi r_x) \cos(\pi r_y)\cos(\pi r_z)$. Also the 2D peculiar case of Fig.~\ref{fig2}c has an analog here
with non-decaying crossover along diagonal directions. 

\subsection{Other crystal symmetries}
Our predictions are robust also in different geometries as for example in the {\it triangular lattice} (instead of cubic).
In this case diagonalization of $H_B$ would be done
through plane waves along momentum directions corresponding to the
correct Bravais lattice.  Since the direct lattice has
proper vectors (in 2D now for simplicity) $\vec{v}_1=\hat{u}_x$ and
$\vec{v}_2=\hat{u}_x/2+\sqrt{3}/2\hat{u}_y$, its Bravais lattice has
vectors $\vec{b}_1=2\pi(\hat{u}_x-\hat{u}_y/\sqrt{3})$ and
$\vec{b}_1=4\pi\hat{u}_y/\sqrt{3}$. The momentum expansion should be
done in this directions and the dispersion relation results
$$\omega_{\vec{k}}=\sqrt{\omega_0^2+8\
g(\sin^2(l_1/2)+\sin^2(l_2/2)+\sin^2(l_3/2))}$$ with $l_1=k_x$,
$l_2=k_x/2+\sqrt{3}k_y/2$ and $l_3=k_x/2-\sqrt{3}k_y/2$.  
\begin{figure}[h!]
\includegraphics[width=1.05\columnwidth]{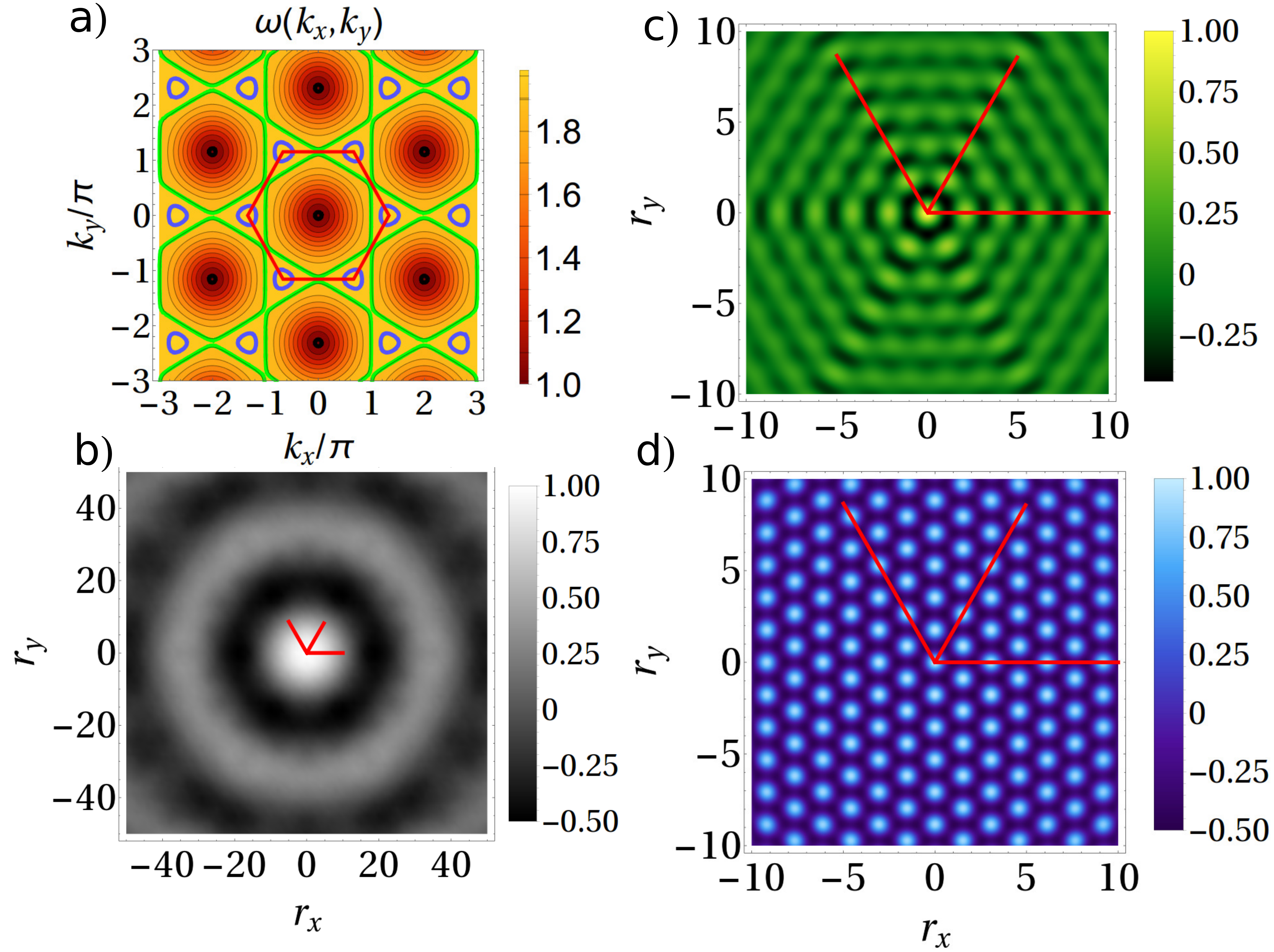}
\caption{a) 2D dispersion relation in colour code with $\omega_0=1$ and $g=0.165$, so that $\omega_{\vec{k}}\in [1,1.992]$. Iso-frequency surfaces are shown
for the limiting cases equivalent to those of the cubic crystal of the main text:  black) $\Omega=1.01$ corresponding to the isotropic case, green) $\Omega=1.905$ directional non-decay, and blue) $\Omega=1.99$ non-decay.
We have also plotted in red the fundamental (Wigner-Seitz) cell, to which momentum integrals are restricted.
Normalized cross-damping term $\Gamma(r_x,r_y)\hat{=}\Gamma_{13}^{(2)}(\vec{r})/\Gamma_{13}^{(2)}(0)$ for b) the isotropic case (low momenta),
for c) directional non-decay (medium momenta) and d) non-decay (high momenta), where we have added in red the crystal symmetry directions to show that the cross-damping term conserves the symmetry of the 
problem. } 
\label{triangular}
\end{figure}
The behaviour of dissipation displays (see Figure~\ref{triangular}) the same regimes of decaying cross-talk for low momenta, and non-decaying cross-talk for higher momenta along
symmetry-favoured directions.

\subsection{Short time behaviour}
\begin{figure}[h!]
\begin{center}
\includegraphics[width=\columnwidth]{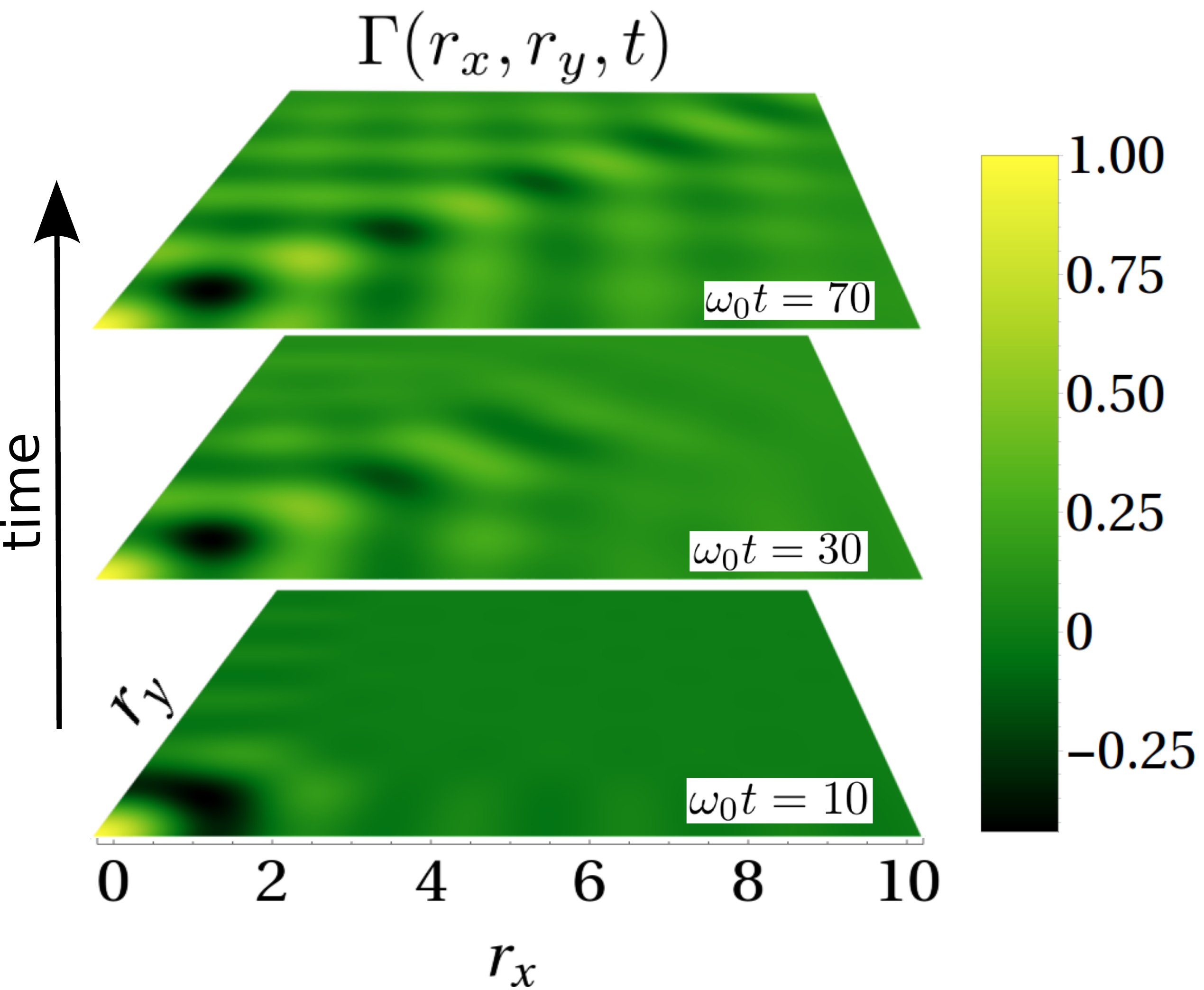}
\caption{Short-time behaviour of the 2D crystal cross-talk, for the case c) of Fig.~\ref{fig2} in main text for times a) $\omega_0t$ = 10,30,70.
The long time limit corresponds to Fig. ~\ref{fig2}c).} 
\label{fig3}
\end{center}
\end{figure}

So far we have discussed the long time limit, relevant for the weak coupling regime, whereas at {\it short times} there is a transient in which the signal 
travels from one probe to the other  at the crystal's fastest group velocity and no cross-damping exist. This is seen in the cross-talk, 
which expands its spatial structure at that velocity (see Fig.~\ref{fig3} and Appendix \ref{AppendixB}), reaching its final (momentum dependent) form (displayed for $t\rightarrow\infty$ in Figs.
\ref{fig1}, \ref{fig2} and \ref{triangular}).

\section{Extended spatial coupling} Considering probes with a {\it finite spatial extension} and hence coupled to a finite-sized region of the crystal, instead of single atoms,  
elucidates the meaning and presence of frequency (momentum) cut-off $\omega_c$ in the description of open systems.  
Even if a crystal presents a natural maximum frequency determined by its periodicity, in open systems the cut-off is often not a property of the environment \cite{Klesse},
depending instead on the probe system. Let us consider probes with extended interaction
$H_{SB}=\lambda\sum_{\vec{R}}g(\vec{R})(q_1
Q_{\vec{n}+\vec{R}}+q_2 Q_{\vec{n}'+\vec{R}})$
with $g(\vec{R})$ a function decaying for $|\vec{R}|>0$ up to each probe size. The new cross-term integrand $\tilde{C}_{13}(\vec{r},\vec{k},t)=C_{13}(\vec{r},\vec{k},t)\Phi(\vec{k})$ 
is modified by  a contact form factor $\Phi(\vec{k})=\sum_{\vec{R},\vec{R}'}g(\vec{R})g(\vec{R}')\cos[\vec{k}(\vec{R}-\vec{R}')]$ and the long times, $T=0$, new expression reads
\begin{equation}
\tilde{\Gamma}^{(D)}_{13}(\vec{r})=\frac{\lambda^2}{2\Omega^2(2\pi)^D}\int\!\!d^D\vec{k}\cos(\vec{k}\vec{r})\, \Phi(\vec{k})\,
\delta(\omega_{\vec{k}}-\Omega)
\end{equation}
where $\Phi(\vec{k})$ limits the maximum effective wavenumbers. For a system-probe coupling  $g(\vec{R})\propto\exp(-|\vec{R}|^2/2\sigma^2)$, the factor
$\Phi(\vec{k})\propto \exp(-|\vec{k}|^2\sigma^2)$ leads to filtered integrals, stemming from the fact that a probe of spatial size $\sigma$ detects an average effect on that area and will be unable
to feel the influence of phonons of shorter wavelengths (higher momentum than $1/\sigma$). In practice, in order to reach the situation in Fig.~\ref{fig2}d each probe needs to have a spatial extent
smaller than the crystal spacing, so that it senses the highest available phonon momenta ($\sigma\to 0$, so $g(\vec{R})=\delta_{\vec{R},\vec{0}}$).    

\begin{figure}[h!]
\begin{center}
\includegraphics[width=0.8\columnwidth]{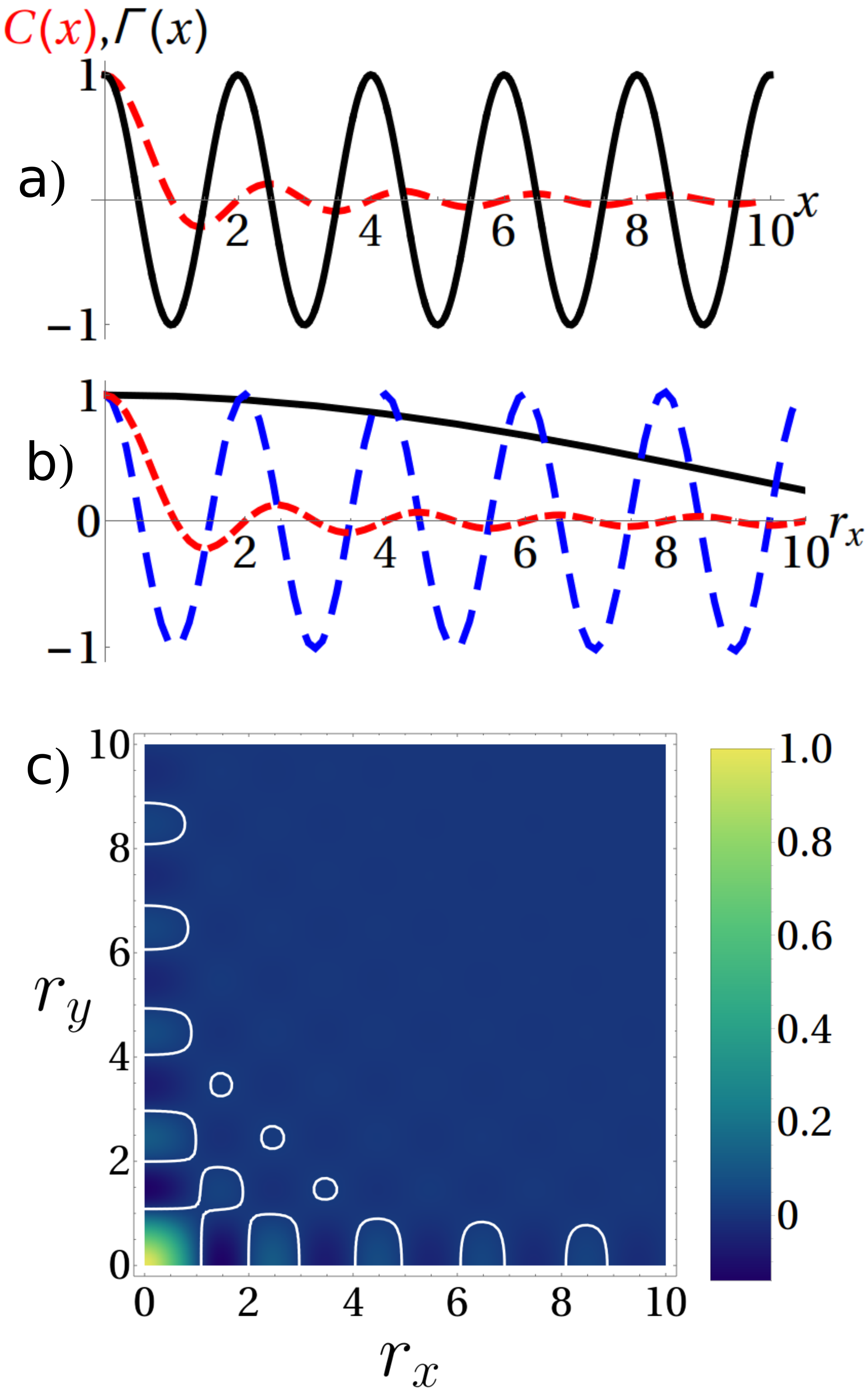}
\caption{We compare here the normalized correlation function $C(r_x,r_y)/C(0,0)$ with the cross-damping in several cases where their
decays do not match at all: a) Crystal correlation function $C(x)$ in 1D in red, vs. the cross-damping term in black of probes with frequency $\Omega=2\omega_0$.
 We have chosen $\omega_0=1$ and $g=3/4\omega_0^2$, so that again $\omega_{\vec{k}}\in [1,2]$. Lower $\Omega$ would simply resonate with a lower
momentum and we would see a cosine with  longer periodicity. b) Correlation function for the 2D-crystal in red, compared with the cross-damping
along $r_x$ (with $r_y=0$) for the isotropic case (black) and high momentum case (blue), as previously shown in Fig.~\ref{fig1}b and \ref{fig1}d, respectively, with the same parameters as figure 1.
c) $C(r_x,r_y)$ in colour code, and we have highlighted the particular value $C(r_x,r_y)=0.01$
in white to guide the eye. This shape does not change significantly for higher temperatures (see Appendix~\ref{AppendixC}). Further, the short range
is not peculiar of this crystal symmetry: a similar behaviour can be observed for the triangular crystal (see Appendix~\ref{AppendixD}).
} 
\label{fig4}
\end{center}
\end{figure}

\section{Correlation length in the crystal} Does the transition from CB to SB we have
seen up to now have to do with the correlation length of the
environment? The quick answer is no, as can be seen in  Fig.~\ref{fig4}a and b. The cross- and self-damping terms
in the dissipation equation (\ref{lindblad}) come from bath operator
spatial correlation functions $\left< Q_{\vec{n}}(0)Q_{\vec{n}'}(t)\right>$ at two times. This time
dependence is the one that, for long times, selects a unique wave vector $\vec{k}_\Omega$
due to resonance with $\Omega$ (through the factor
$\sinc[(\omega_{\vec{k}}-\Omega)t]$) and therefore follows from a reduced manifold $D-1$ of momenta.  On the other hand, the
correlation in the crystal at two different points comes from functions
at equal time $\left< Q_{\vec{n}}(t)Q_{\vec{n}'}(t)\right>$ and
follows from all phonons momenta. In other words cross-damping is caused by  
{\it resonant} phonons, while generic correlations in the
crystal are caused by interference of {\it all} phonons thus decaying with
distance even in 1D (Fig.~\ref{fig4}a). 
Usually, as in our case, the bath is in a stationary (thermal) state, and thus the correlation function is time-independent
$C(r_x,r_y)=\left<Q_{\vec{0}}\, Q_{\vec{r}}\right>$, with  
 $\left<Q_{\vec{r}}\, Q_{\vec{r}+\vec{R}}\right>=(2\pi)^{-D}
\int_{-\pi}^\pi d^D\vec{k} \cos(\vec{k}\cdot\vec{R})\left[n(\vec{k})+1/2\right]/\omega_{\vec{k}}$.
The 2D case (Fig.~\ref{fig4}b and c) clearly displays spatial correlations decay at distances of 
the order of the crystal lattice $\xi_c\approx a$ (notice that all spatial coordinates are scaled with
$a$ in the rest of the manuscript), being stronger along crystal directions, while the cross-talk decays on a scale given by the resonant 
normal mode wave-length $\approx |\vec{k}_\Omega|^{-1} $ (isotropic case) or does not decay at all (anisotropic case).

\section{Discussion}

\subsection{Generality}Our conclusions can be generalized to other system-bath models, e.g. where the environment is a non-interacting field which exchanges excitations with 
the system probes: i.e a collection of free particles with a given dispersion relation $\epsilon(\vec{k})$ whose eigenfunctions have a spatial
profile $f(\vec{r},\vec{k})$, so that the bath Hamiltonian is $H_B=\sum_{\vec{k}}\epsilon_{\vec{k}}b_{\vec{k}}^\dagger b_{\vec{k}}$ 
(or $\int d^D\vec{k} \epsilon_{\vec{k}}b_{\vec{k}}^\dagger b_{\vec{k}}$ for continuous spectra); the exchange interaction between
bath and system probes being $H_{SB}\propto\sum_{\vec{k}}[f(\vec{r},\vec{k})a_1 +f(\vec{r}+\vec{R},\vec{k})a_2]b_{\vec{k}}^\dagger+h.c.$, so probe 1 is located at $\vec{r}$
and probe 2 at $\vec{r}+\vec{R}$.
In such case and assuming secular and Born-Markov regime, the cross-talk is given by
\begin{equation}
\label{cross}
\Gamma_{13}^{(D)}(\vec{r},\vec{r}+\vec{R})\propto\int d^D\vec{k}f(\vec{r},\vec{k})
f^*(\vec{r}+\vec{R},\vec{k}) \delta(\epsilon_{\vec{k}}-\Omega)g(\epsilon_{\vec{k}})
\end{equation}
where the function $g(\cdot)$ is related to
how the probes couple to each mode.
The free particles could be Bogoliubov bosons on top of a condensate in an optical lattice, electrons in the bulk,
phonons in a crystal with disorder (as in the main text) or
any other free particles which, because of the locality and weakness of the probe-bath coupling lead to such master equation
with this cross-term. The delta function is a consequence of the fact that the system-bath is energy exchanging, and thus that
we have {\it dissipation}. Note also that we have assumed that different bath modes are uncorrelated and stationary, as usual in e.g. a
thermal state.

The bath free field can be expressed in terms of the single-particle operators, leading to a correlation
function 
\begin{equation}
\label{correl}
\left< \phi(\vec{r},t)\phi(\vec{r}+\vec{R},t)\right>\propto\int d^D\vec{k}f(\vec{r},\vec{k})
f^*(\vec{r}+\vec{R},\vec{k})h(\epsilon_{\vec{k}})
\end{equation}
(notice that there are two generic functions $h$ and $g$ which are model-dependent).
For non-interacting fields it is thus clear that CB/SB cross-over distance is unrelated to the correlation length in the medium, simply because
the former is propagated by resonant free particles, while the latter is propagated by all particles.
Other models with more complicated interactions than just particle exchange, or even interacting models for the bath, might
yield different behaviours and are subject of future interest.

We comment now on how the phenomenology studied for the D-dimensional crystal translates into this generic class of models:
a) in the 1D case the cross-damping will be of the form $\Gamma_{13}^{(1)}(\vec{r},\vec{r}+\vec{R})\propto f(\vec{r},\vec{k}_\Omega)
f^*(\vec{r}+\vec{R},\vec{k}_\Omega)g(\epsilon_{\vec{k}_\Omega})$, meaning that the overlap of mode functions (of eigenmode $\vec{k}_\Omega$)
between the two probe positions will dictate the decay from CB to SB, i.e. the spatial shape of $f(\vec{r},\vec{k}_\Omega)$, be it localized or periodic,
will lead to decay or non-decay respectively; b) the very peculiar behaviour observed in Fig.~\ref{fig2}c,d requires very well-matched interference of
plane waves (thus a translational invariant medium) and thus is not to be expected in general; c) the short/long time argument is based on the 
nature of the {\it sinc} function and thus independent on the details of the model, hence any possible long-range cross-damping will take a time to build up,
related with the fastest excitations in the environment; finally, d) also irrespective of the details of the bath model, the presence of a probe with finite spatial extension
will blur any short-range (high momentum) details, leading to a high-momentum cutoff in the integrals defining the coefficients of the master equation.

\subsection{Dephasing} A further interesting point is to consider the comparison with the case of a {\it dephasing model}.
In that situation the cross-talk will {\it not} have any resonance constraint imposed, and thus the two integrals eqs. (\ref{cross}) and (\ref{correl}) will
be similar except for the functions $h(\epsilon_{\vec{k}})$ and $g(\epsilon_{\vec{k}})$, leading to similar behaviours.
Some typical bath's spectral densities (encoding function $g(\cdot)$ and the density of states of the bath)
$\omega^D \exp(-\omega/\omega_c)$ favour small momenta in the cross-talk for 1D, while for 3D they favour
frequencies/momenta near the cut-off frequency $\omega_c$ \cite{palma,kohler2006}. Thus for pure dephasing
the CB to SB transition length will be similar to the correlation length of the environment.

\subsection{Experimental implementations}

One possible way to experimentally implement the
2D crystal is via trapped ions with a tight axial confinement so that
they effectively lie on a plane and form a triangular-symmetric Coulomb
crystal, such as in \cite{porras}.  The major problem in that setting is
that axial motion is coupled to radial degrees of motion, but this can
be overcome if the axial frequency is sufficiently higher than the
radial counterpart. The probe ions would need to be sitting in the same
plane thus distorting the modes of the Coulomb crystal. Therefore the
modelling would be slightly more complicated , although the basic
physics would be the same. Addressability of the probe ions, e.g. by
fluorescence \cite{wine2003}, would be a central requirement.

Another possible way of investigation is the intentional deposition of
atoms adsorbed in metallic surfaces. This has always been considered as
a drawback and a source of anomalous heating in ion trap electrodes
\cite{wineNoise,blattNoise}, but could suit our purposes. Adsorbed atoms
bound to a metallic surface can have oscillation frequencies in the THz
regime, very close to Debye frequencies of metals (gold for example has
a Debye frequency of around 3.6 THz). In this way, by placing
intentionally adsorbed atoms at different distances would allow us to
check our results. Different masses of these atoms would scan the
different frequencies as compared to the maximum phonon frequency of the
metallic substrate. For this to be possible we should deal with
fluorescent adatoms which can be addressed and localized by lasers.
Investigation of cross-damping could be done by exciting the motion of
one atom and evaluating the effect on the other. A coupling of the
fluorescent transition to the motional degree of freedom would probably
be needed, though.

\subsection{Outlook}
An immediate consequence of this work is that initial correlations between two dissipating units
will be highly sensitive to details of the underlying medium, such as crystal symmetries. This suggests a possible avenue 
to use multi-party quantum systems to test/probe media with unknown properties. One could further 
envision the use of a lattice of coupled probes to obtain information of an unknown surface through
the decay of spatial modes of the probe-lattice. In this direction, recent work \cite{haffner} has 
shown that a single trapped ion can be confined near a metallic surface to extract electric-field
noise characteristics through its heating rate.
Also, in view of recent proposals to use surface acoustic waves as a quantum bus between many different types
of quantum systems \cite{ciracPRX}, the phenomenon of preferential directions seen in figures~\ref{fig2} and \ref{fig3}
could be potentially used to build substrates with a patterned surface whose symmetry allows for distant units
to communicate along diagonal/triangular directions with a decay only given by static imperfection (disorder)
of the material. 
All these avenues are left for future investigation.

\section{Conclusion} Do the separate units of a spatially extended system suffer dissipation and decoherence from common or separate baths?
We tackled this fundamental issue introducing a microscopic environment model where spatial distances and correlations appear naturally. 
Beside the ineffectiveness of environment spatial correlations to determine this transition, we have shown 
the importance of dimensionality, symmetries and probes extensions. The prediction of collective dissipation between distant probes in a 1D 
homogeneous environment when placed at a distance multiple of $2\pi k_\Omega^{-1}$ opens up interesting possibilities  in  surface phononic cavities 
\cite{ciracPRX} and phonon wave-guides \cite{zollernjp}. Similar predictions can hold for planar or bulk platforms environments,
for probes at relative position now determined both by their oscillations frequency and the crystal symmetries. Indeed when $D>1$, the dispersion 
is isotropic for $\Omega \ll c/ \ell$, with $c$ the effective propagation velocity in
the medium and $\ell$  either wavelength of the crystal periodicity or the mean distance between disorder patches in an
otherwise homogeneous medium \cite{blattNoise}. The anisotropy opens a communication channel (resulting from the interference of a manifold of resonant phonons)
between the probes, even at large distance while the effect is degraded in presence of disorder. 
On the other hand, independent dissipation will occur for rather distant  and `slowly oscillating' probes, when the effective dispersion is 
isotropic as in the largely studied case of electromagnetic fields in homogeneous media. 

Collective and local dissipation of multipartite systems in crystal environments can be extended 
to frontline platforms that can serve as substrates in quantum technologies, such as  metamaterials with gapped 
spectra or displaying topological modes  \cite{kane,palouse}, and in polaritons configurations \cite{plenio}, 
optomechanical arrays \cite{xuereb,marquardt} or cold atoms in different phases \cite{Greiner}. Furthermore, even if disorder 
in 1D environments has been shown to hinder collective dissipation, there are several open questions in larger 
dimensions and in presence of phenomena such as Anderson localization \cite{AL}. 

This work has been supported by EU through the H2020
Project QuProCS (Grant Agreement 641277), by projects NoMaQ FIS201460343-P and QuStruct FIS2015-66860-P (MINECO/FEDER), and by UniMI through 
the H2020 Transition Grant 15-6-3008000-625. AM acknowledeges support by EU-LLP Erasmus placement 
program, FG from UIB's postdoctoral program.

\appendix

\section{Master equation in periodic and disordered environment}
\label{AppendixA}

We consider a $D$-dimensional harmonic crystal with nearest neighbour interactions:
$H_B=\sum_{\vec{n}} \frac{P_{\vec{n}}^2}{2} +
\frac{\omega_0^2 Q_{\vec{n}}^2}{2} + \frac{g}{2}
\sum_{\vec{l}} (Q_{\vec{n}}-Q_{\vec{n}+\vec{l}})^2$
where $\vec{n}\equiv (n_1,n_2,...n_d)$ is the site index where 
each mass lies, and $\vec{l}$ are unit lattice vectors, being  for a cubic structure $\vec{l}\in\{\hat{u}_x,\hat{u}_y,....,\hat{u}_D\}$ 
in each of the $D$ spatial directions. The probes are first considered as point-like coupled to the bath
at points $\vec{n}$ and $\vec{n'}=\vec{n}+\vec{r}$, so the system-bath interaction is $H_{SB}=\lambda(q_1 Q_{\vec{n}}+q_2 Q_{\vec{n}'})$
The overall Hamiltonian is given by $H=H_B+H_S+H_{SB}$
where the extended system consists of the two identical uncoupled $H_S=\Omega(a_1^\dagger a_1+a_2^\dagger a_2)$ harmonic probes.
Notice that we introduce only one degree of freedom for each site, which corresponds also
to a model of a scalar field with spatial discrete structure. If we set $\omega_0=0$, in 3D it also can be associated to studying
cross-talk mediated by phonons of only one polarization in a realistic crystal, as for example gold \cite{crystals}, 
with a linear anisotropic dispersion that saturates for high momenta. Since dissipation into the crystal can always be decomposed
into three polarizations, we can choose to match the probe-to-probe direction, thus separating the problem into the three sets of
polarizations, each having an anisotropic dispersion relation, as here considered.
%
%

The master equation of the system (two probes) density matrix up to the second order in the coupling strength, is obtained in the Born-Markov
approximation \cite{breuer} and given by
\begin{equation}
\dot{ \tilde{\rho}}_S(t) = -\int_0^t d\tau \Tr_{B} \{ [ \tilde{H}_{SB}(t), [\tilde{H}_{SB}(t-\tau), R_0 \otimes  \tilde{\rho}_S(t) ]] \} 
\end{equation}
in the interaction picture $\tilde{\rho}_S$, where $\tau = t' - t$ and $R_0=\exp(-\beta H_B)/Z_B$ the invariant thermal state of the (crystal) environment.
In the crystalline case, the bath Hamiltonian is diagonalized by plane waves, and the system-bath Hamiltonian is then

\begin{equation}
  H_{SB}= \int d^D\vec{k}\frac{\lambda}{2\sqrt{\Omega\ \omega_{\vec{k}}}} (S^\dag_{\vec{k}}A_{\vec{k}} + S_{\vec{k}}A_{\vec{k}}^\dag  ),
\end{equation}
with $S_{\vec{k}}=(2\pi)^{-D/2}(a_1 e^{i\vec{k}\vec{n}} +a_2 e^{i\vec{k}\vec{n}'})$.
In the case of a crystal with disorder, translational invariance is broken and the bath is not any more diagonalized by plane waves, 
but by the general transformation $Q_{\vec{n}}=\int_{-\pi}^{\pi}d\vec{k}\ f_{\vec{n},\vec{k}} Q_{\vec{k}}$
and the system operators read $S_{\vec{k}}=a_1 f_{\vec{n},\vec{k}}+a_2 f_{\vec{n}',\vec{k}}$.

After some standard algebraic operations, and going back to Schr\"odinger picture, the master equation reduces to
\begin{widetext}
\begin{eqnarray}
\dot{\rho}_S (t)&=&-i[H_S+H_{LS},\rho_S (t)]+\frac{\lambda^2}{2\Omega}\int_{-\pi}^{\pi}d^D\vec{k}\ 
 \frac{1}{\omega_{\vec{k}}}\frac{\sin[t(\Omega-\omega_{\vec{k}})]}{\Omega-\omega_{\vec{k}}}\left\{ ( N_{\vec{k}}+1){\ele}_{S_{\vec{k}}}(\rho_S)+ N_{\vec{k}}{\ele}_{S_{\vec{k}}^\dagger}(\rho_S)\right\} \nonumber \\
{\ele}_{O} (\rho_S)&=& O\rho_S O^\dagger -\frac{1}{2}\{ O^\dag\,  O,\rho_S \}\nonumber\\
H_{LS}&=&\frac{\lambda^2}{2\Omega}\int_{-\pi}^{\pi}d^D\vec{k}\ \frac{1}{\omega_{\vec{k}}}\frac{1-\cos[t(\Omega-\omega_{\vec{k}})]}{\Omega-\omega_{\vec{k}}} S_{\vec{k}}^\dagger S_{\vec{k}} \label{fullME}
\end{eqnarray}

with $N_{\vec{k}}=\langle A_{\vec{k}}^\dag A_{\vec{k}}\rangle_{_{R_0}}$, and substituting the corresponding operators $S_{\vec{k}}$ in the equations.
In terms of $F_i = \{ a_1, a_1^\dag, a_2, a_2^\dag \}$, the dissipative part reads $\dot{ \rho}_S =  \sum_{j ,l=1}^4 \Gamma_{j l}(\vec{r},t) ( F_j \rho_S F_l^\dag - \frac{1}{2} \{F_l^\dag F_j,\rho_S\})$, 
with
%
\begin{equation}
   \begin{split}
 \Gamma_{11}(t)& = \Gamma_{33} (t)= \frac{\lambda^2}{2\Omega}\int_{-\pi}^{\pi}d^D\vec{k}\ \frac{N_{\vec{k}}+1}{\omega_{\vec{k}}}\ \frac{\sin[t(\Omega-\omega_{\vec{k}})]}{\Omega-\omega_{\vec{k}}}|f_{\vec{n},\vec{k}}|^2 \\ 
 \Gamma_{22}(\vec{r}, t) &= \Gamma_{44}(\vec{r}, t)= \frac{\lambda^2}{2\Omega}\int_{-\pi}^{\pi}d^D\vec{k}\ \frac{N_{\vec{k}}}{\omega_{\vec{k}}}\ \frac{\sin[t(\Omega-\omega_{\vec{k}})]}{\Omega-\omega_{\vec{k}}}|f_{\vec{n},\vec{k}}|^2 \\
\Gamma_{13} (\vec{r}, t)&=\Gamma_{31} (\vec{r}, t)=\frac{\lambda^2}{2\Omega}\int_{-\pi}^{\pi}d^D\vec{k}\ \frac{N_{\vec{k}}+1}{\omega_{\vec{k}}}\ \frac{\sin[t(\Omega-\omega_{\vec{k}})]}{\Omega-\omega_{\vec{k}}}f_{\vec{n},\vec{k}}f_{\vec{n}',\vec{k}}^* \\ 
\Gamma_{24}(\vec{r},t) &=\Gamma_{42}(\vec{r},t)=\frac{\lambda^2}{2\Omega}\int_{-\pi}^{\pi}d^D\vec{k}\ \frac{N_{\vec{k}}}{\omega_{\vec{k}}}\ \frac{\sin[t(\Omega-\omega_{\vec{k}})]}{\Omega-\omega_{\vec{k}}}f_{\vec{n},\vec{k}}f_{\vec{n}',\vec{k}}^*, \\
   \end{split}
\end{equation}
always understanding that $\vec{r}=\vec{n}-\vec{n}'$. Correspondingly :
\begin{eqnarray}
\Delta\Omega&=&\frac{-\lambda^2}{4\Omega}\int_{-\pi}^{\pi}d^D\vec{k}\ \frac{1-\cos[t(\Omega-\omega_{\vec{k}})]}{\omega_{\vec{k}}(\Omega-\omega_{\vec{k}})}|f_{\vec{n},\vec{k}}|^2\\
\gamma&=&\frac{-\lambda^2}{4\Omega}\int_{-\pi}^{\pi}d^D\vec{k}\ \frac{1-\cos[t(\Omega-\omega_{\vec{k}})]}{\omega_{\vec{k}}(\Omega-\omega_{\vec{k}})}f_{\vec{n},\vec{k}}f_{\vec{n}',\vec{k}}^*.
\end{eqnarray}

In the crystalline case (no static disorder), we have $f_{\vec{n},\vec{k}}\propto e^{i\vec{k}\cdot\vec{n}}$, so $|f_{\vec{n},\vec{k}}|^2=1$ and $f_{\vec{n},\vec{k}}f_{\vec{n}',\vec{k}}^*=e^{i\vec{k}\cdot\vec{r}}$ (which leads, throught the symmetry
of $\omega_{\vec{k}}$ to $\cos(\vec{k}\cdot\vec{r})$ in the main text).

\newpage

\section{Short time behaviour}
\label{AppendixB}

\begin{figure}[hb!]
\includegraphics[width=0.82\textwidth]{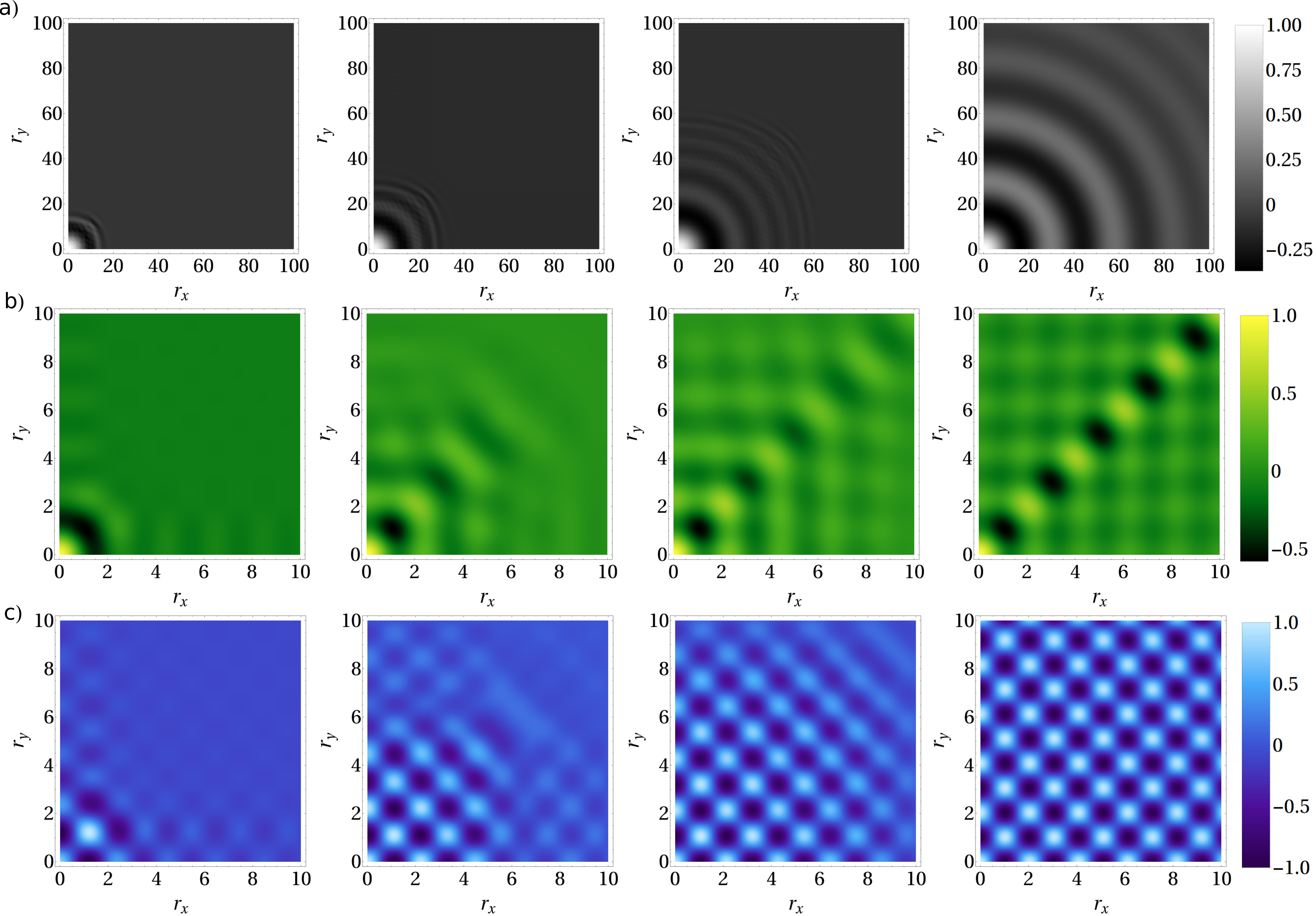}
\caption{Short-time behaviour of the 2D crystal cross-talk, for the limiting cases of figure 2 in main text: a) isotropic, b) directional non-decay, c) non-decay.
From left to right, we show times a) $\omega_0 t=50,100,200,1000$ and b,c) $\omega_0 t=10,30,70,10000$. For b) and c) it was necessary to plot longer times in order to see
better the resulting cross-talk obtained in main text's figure 2 for the long time limit.} 
\label{supp3}
\end{figure}

\end{widetext}



\section{Correlation function for finite temperature}
\label{AppendixC}
The environment's correlation function at finite temperature is 
\begin{eqnarray}
&\left<\phi(\vec{r})\phi(\vec{r}+\vec{R})\right>\propto \int_0^\pi d^D\vec{k} \left[2n(\vec{k})+1\right] \cos(\vec{k}\cdot\vec{R})&\nonumber\\
&=\int_0^\pi d^D\vec{k} \coth\left[\frac{\hbar\omega_{\vec{k}}}{2k_BT}\right] \cos(\vec{k}\cdot\vec{R})&\nonumber
\end{eqnarray}
, and therefore the temperature dependence contributes to its spatial shape. This is in contrast with the long-times behaviour of the cross-talk at finite temperature 

\begin{figure}[h!]
\includegraphics[width=\columnwidth]{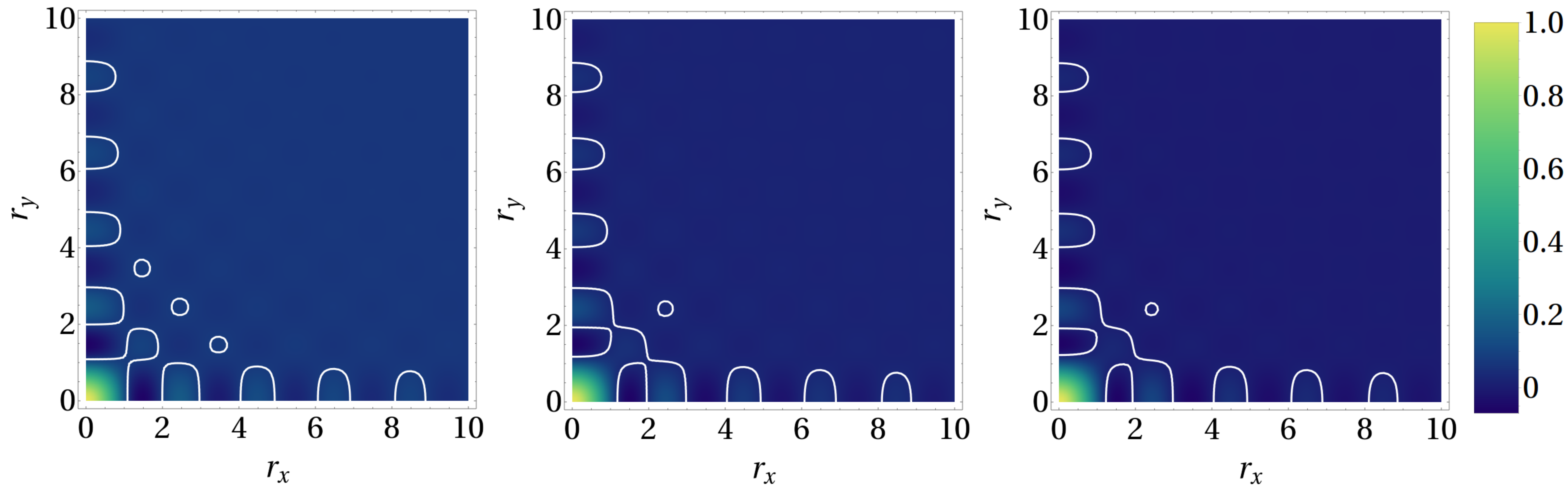}
\caption{Normalized correlation function at finite temperature $C(\vec{r},T)$ for the 2D cubic crystal in colour code. From left to right, $T/\omega_0=0,1,100$. We stress the prominently 
weak influence of temperature on most features, specially its spatial distribution.} 
\label{supp4}
\end{figure}

\begin{eqnarray}
&\Gamma_{13}^{(D)}(\vec{r})\propto\int_{-\pi}^\pi d^D\vec{k}\delta(\omega_{\vec{k}}-\Omega)\left[n(\vec{k})+1\right]\cos(\vec{k}\cdot\vec{R})& \nonumber\\
&=\int_{-\pi}^\pi d^D\vec{k}\delta(\omega_{\vec{k}}-\Omega)\frac{1}{2}\left[\coth\left(\frac{\hbar\omega_{\vec{k}}}{2k_BT}\right)+1\right]\cos(\vec{k}\cdot\vec{R})&\nonumber
\end{eqnarray}
where the cotangent factors out of the integral. Thus we have a common prefactor $(1/2)\left[\coth(\hbar\Omega/2k_BT)+1\right]$ and an integral in momenta which does not depend on temperature, meaning that the spatial shape is independent
of temperature.
It must be stressed though, as seen in Fig.~\ref{supp4}, that the correlation function is not too different for different temperature scales, mostly its basic spacing which coincides with the
crystal constant.

\section{Correlation function with triangular symmetry}
\label{AppendixD}

The correlation function respects the symmetric directions of the crystal but also decays fast (on the order of the crystal spacing) as in the case of cubic symmetry.

\begin{figure}[h!]
\begin{center}
\includegraphics[width=0.8\columnwidth]{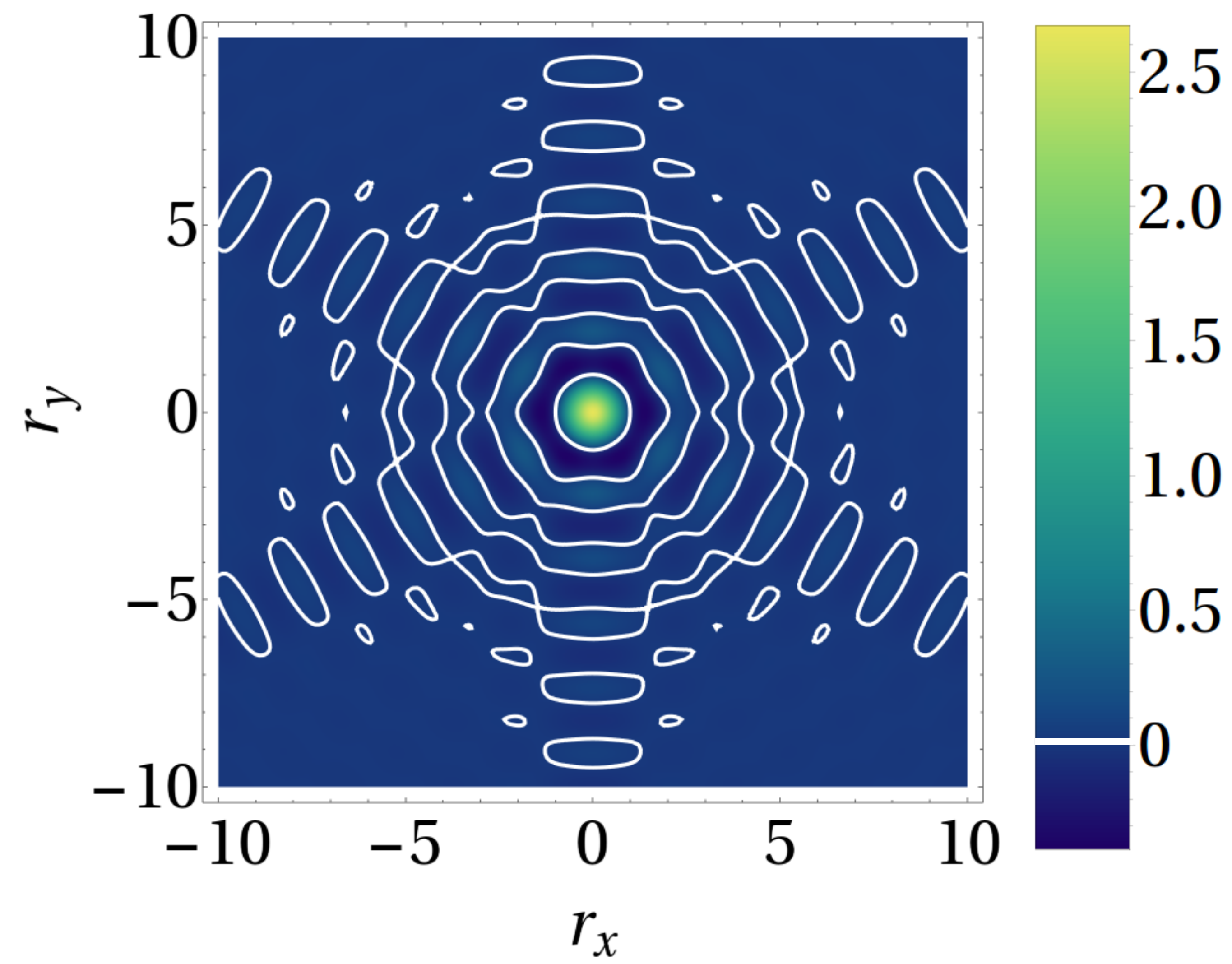}
\caption{Correlation function $C(r_x,r_y)$ for the 2D triangular crystal in colour code. We have highlighted the particular value $C(r_x,r_y)=0.01$
in white to guide the eye. } 
\end{center}
\label{supp2}
\end{figure}

\end{document}